\documentstyle[prl,aps,preprint]{revtex}
\begin{document}
\tighten
\title{DOES THE GLUON SPIN CONTRIBUTE IN A\\
   GAUGE-INVARIANT WAY TO NUCLEON SPIN?} 

\author{Pervez Hoodbhoy\thanks{Permanent Address:
Department of Physics, Quaid-e-Azam University, 
Islamabad 45320, Pakistan} and Xiangdong Ji}
\address{Department of Physics \\
University of Maryland \\
College Park, Maryland 20742 }
\vspace{0.2in}
\date{UMD PP\#00-008 ~~~DOE/ER/40762-190~~~ August 1999}
\maketitle

\begin{abstract}

Although the matrix element of the gluon spin operator in nucleon 
helicity states is known to be independent of some special choices 
of gauge, we show that it is not invariant under a general gauge 
transformation. We find that there exists a simple means of 
obtaining the matrix element in a different choice of gauge 
from a calculation made in one specific
gauge. Similar conclusions hold for other manifestly gauge-dependent
operators present in the QCD angular momentum operator.

\end{abstract}
\pacs{xxxxxx}

\narrowtext

In quantum chromodynamics (QCD), the gluon spin 
operator, $\vec{S}_g$, can be defined as,
\begin{equation}
   \vec{S}_g = \int d^3x~ \vec{E}\times \vec{A} \ , 
\end{equation}
where we have suppressed the color indices, 
$\vec{E}$ is the color electric field and 
$\vec{A}$ the gauge potential. The above 
operator is manifestly gauge dependent, and 
therefore one expects that its matrix 
element in a physical state 
depends on gauge choice. Indeed, in Ref. 
\cite{hoodbhoy} it was shown that the 
matrix element of $\vec{S}_g$ in a quark 
helicity state is different in 
the axial ($A^+=0$) and covariant gauges. 
On the other hand, it was claimed
in Ref.\cite{chen} that this matrix element in
a helicity state is invariant under a gauge 
transformation. The goal of this paper
is to clarify the relationship of the two results. 
In particular, we point out that the proof in
Ref.\cite{chen} covers only a special class of
gauge transformations and the matrix element of 
$\vec{S}_g$ {\it is} in fact gauge dependent.
We then make a number of pertinent 
observations about comparing calculations in 
different gauges. In particular, we demonstrate
that there exists a simple means of getting
the answer in a different choice of
gauge from calculations made in 
one specific gauge.
 
We start with general comments 
about calculations in quantum gauge theories. 
Because of the gauge symmetry and the infinite 
number of degrees of freedom in fields, it is 
always necessary to choose a gauge for calculating
Green's functions and physical matrix elements, 
perhaps with the exception of lattice gauge
theory in which fields are assigned at a finite
number of spacetime points. Indeed, in  a
canonical quantization of gauge theories,
one has to choose a gauge at the very outset. 
In the path integral
formulation, a gauge choice is conveniently made
by selecting a set of gauge conditions,
\begin{equation}
      F^a(A)= w^a(x) \ , 
\end{equation}
where the index $a$ runs over the number of generators
of the gauge group (and will be omitted in the
remaining equations). In addition, one can choose
an arbitrary weighting functional 
$G(w)$ to integrate over the auxiliary field $w$. 
Without loss of generality, one can assume
$\int [Dw]G(w)=1$, where $D$ denotes the functional
integration measure. Hence, the generating 
functional for Green's functions in QCD
reads \cite{itzykson},
\begin{equation}
   Z(J) = {1\over \cal{N}}\int [D\phi][Dw] \Delta_F(A)
         \delta(F(A)-w)G(w) \exp\left(iS 
      + i\int d^4x [J\cdot A+\bar\eta\psi
    + \bar\psi\eta]\right) \ , 
\end{equation}
where $\phi$ denotes a collection of fields $A$, $\psi$, 
and $\bar \psi$, and $S$ is the canonical QCD action, 
\begin{equation}
S = \int d^4x
\left[\overline{\psi}(i\!\not\!\! D-m_q)\psi - 
  {1\over 4}F^{\mu\nu} F_{\mu\nu} \right] \ , 
\end{equation}
and the normalization constant ${\cal N}$ is
\begin{equation}
    {\cal N} = \int [D\phi][Dw] \Delta_F(A)
         \delta(F(A)-w)G(w) e^{iS} \ , 
\end{equation}
so that $Z[0] = 1$. The Faddeev-Popov determinant
$\Delta_F(A)$ is defined such that
\begin{equation}
     \Delta_F(A)\int [D\omega] \delta(F(^\omega A)-w) = 1
\end{equation}
where $^\omega A$ is a gauge transformation of $A$ and
$[D\omega]$ integrates over the gauge group at every
point of spacetime.

To calculate the physical matrix element $\langle
f|\hat O(\phi)|i\rangle$ of an operator $\hat O(\phi)$, 
one can start with the following Green's function,
\begin{eqnarray}
 && \langle 0|
T(J_f(x)O(y)J_i^\dagger(0))|0\rangle|_{F,G} \nonumber \\
 &=& {1\over {\cal N}}\int [D\phi][Dw] \Delta_F(A)
         \delta(F(A)-w)G(w) J_f(\phi)
        O(\phi) J_i^\dagger(\phi) e^{iS}\ ,  
\label{matrix1}
\end{eqnarray}
where $J_{i,f}$ are the interpolation fields
or currents for the initial and final physical states. 
According to the Lehmann-Symanzik-Zimmermann
reduction formula \cite{itzykson}, $\langle
f|\hat O(\phi)|i\rangle$ is just the residue
of the Green's function at the poles
corresponding to external physical states, 
modulo the coupling constants
$Z_{i,f}$ defined as 
$\langle 0|J_{i,f}(0)|i,f\rangle 
= \sqrt{Z_{i,f}}$ (where a possible 
Lorentz structure is suppressed).
Gauge invariance of the physical matrix element
means that the residue depends on a
choice of $F(A)$ and $G(w)$ only through
the couplings $Z_{i,f}$ when the interpolating 
currents are gauge dependent.

Using methods available in textbooks\cite{itzykson},
one can show that the Green's function 
$\langle 0|
T(J_f(x)O(y)J_i^\dagger(0))|0\rangle|_{F,G}$ 
is independent of the choice of $F(A)$ and $G(\phi)$
provided that
$J_i(\phi)$, $J_f(\phi)$, and $O(\phi)$ are
gauge invariant. Indeed, consider a different choice 
of $\tilde F$ and $\tilde G$ such that
\begin{equation}
  \Delta_{\tilde F}(A)\int [D\omega]\delta(\tilde F(^\omega A)-\tilde w) = 1; 
   ~~~~\int [D\tilde w]\tilde 
    G(\tilde w) = 1
\end{equation}
Multiplying both unities to the right-hand-side 
of Eq. (\ref{matrix1}), we get
\begin{eqnarray}
&&  \langle 0|
T(J_f(x)O(y)J_i^\dagger(0))|0\rangle|_{F,G} \nonumber  \\
 &=& {1\over {\cal N}}\int [D\phi][Dw][D\tilde w]
          [D\omega] ~\Delta_F(A)
         \delta(F(A)-w)G(w) J_f(\phi)
        O(\phi) J_i^\dagger(\phi)  \nonumber \\
    && \times \Delta_{\tilde F}(A) \tilde G(\tilde w) 
     \delta(\tilde F(^\omega A)-\tilde w) e^{iS} \nonumber \\
  &=& {1\over {\cal N}} \int [D\phi][Dw][D\tilde w]
          [D\omega] ~\Delta_F(A)
         \delta(F(^{\omega^{-1}}A)-w)G(w) J_f (^{\omega^{-1}}\phi)
        O(^{\omega^{-1}}\phi)  \nonumber \\
  &&  \times J_i^\dagger(^{\omega^{-1}}\phi) \Delta_{\tilde F}(A)
     \tilde G(\tilde w)  \delta(\tilde F(A)-\tilde w)e^{iS} \nonumber \\
 &=&  {1\over {\cal N}}\int [D\phi][Dw][D\tilde w]
       ~G(w) J_f (^{\omega_0}\phi)
        O(^{\omega_0}\phi) J_i^\dagger (^{\omega_0}\phi) 
     \Delta_{\tilde F}(A) \tilde G(\tilde w) 
     \delta(\tilde F(A)-\tilde w) e^{iS}  \nonumber \\
 & = & {1\over {\cal N}} \int [D\phi][D\tilde w]
       ~J_f(\phi)
        O(\phi) J_i^\dagger(\phi) 
     \Delta_{\tilde F}(A) \tilde G(\tilde w) 
     \delta(\tilde F(A)-\tilde w)e^{iS} , 
\end{eqnarray}
where after the second equal sign, we have made 
the gauge transformation $\phi\rightarrow ^{\omega^{-1}}\phi$ 
and used the fact that the measure
$[D\phi]$ and determinants $\Delta_F(A)$ 
and $\Delta_{\tilde F}(A)$ 
are invariant under the transformation. After the
third equal sign, we have integrated out the $[D\omega]$
by using the $\delta$-function constraint 
$F(^{\omega^{-1}}A)=w$. 
For a fixed $A$ and $w$, there is a special $\omega_0$
that fulfills the constraint, and hence $\omega_0$
is a specific function of them: $\omega_0 = \omega^{-1}(A,w)$. 
After the fourth equal sign, we have used the 
assumption that $J_{i,f}(\phi)$ and $O(\phi)$ are 
gauge invariant. This makes the functional integration
over $w$ trivial. The final line establishes
gauge independence of the Green's function:
$ \langle 0|
T(J_f(x)O(y)J_i^\dagger(0))|0\rangle|_{F, G}
=  \langle 0|
T(J_f(x)O(y)J_i^\dagger (0))|0\rangle|_{\tilde F,\tilde G}$. 

From the above discussion, we see  
that gauge invariance of a matrix 
element is guaranteed when the observable
under consideration, defined as a functional 
of physical fields in the
path integral formalism, is invariant under {\it any}
gauge transformation $\omega$, including 
arbitrary dependence on the gauge 
potential $A$ itself. The notion
of $A$-dependent gauge transformation can be 
found in textbooks, but it is usually discussed 
in the context of classical
gauge theory. In a canonically-quantized theory,
an $A$-dependent $\omega$ can no longer
be considered as a transformation ``parameter'', 
but rather a quantum operator in Hilbert space.
Because the commutation relations between 
$\omega$ and the fundamental fields depend
on the choice of transformation itself,
it is difficult to consider the most general 
gauge transformations in canonically-quantized 
theories. In fact, as we have alluded to before, 
canonical quantization of a gauge theory can only be 
carried out after a choice of gauge,
and different gauge choices often lead 
to different quantum Hilbert spaces. When comparing 
calculations in different gauges, one usually
compares the final results only$-$not the actual 
physical states and operators in those gauges.
In particular, we know of no discussion in 
the literature about how to transform a quantum 
operator from one gauge to another. 

Fortunately, in the path-integral formulation, 
all field variables are treated as classical,
and a transformation from one gauge to another, 
performed with an $A$-dependent $\omega$, 
can be easily handled by a change of integration 
measure and the associated Jacobian (the Faddeev-Popov 
determinant). A decisive test of gauge invariance of an observable
can be done with a general gauge 
transformation: an $\omega(A)$ solving
$F(^\omega A)=w$ for any $A$ satisfying 
$\tilde F(A)=\tilde w$. It can happen that 
certain matrix elements are 
invariant under a special class of gauge 
transformations; in particular, those that keep
the Faddeev-Popov determinant invariant. However,
if they cannot pass the general test above, one
cannot claim their total gauge independence. 

In Ref. \cite{chen}, Chen and Wang claim to have 
shown that the matrix element of $\vec{S}_g$
in a helicity eigenstate is gauge invariant. 
What they have actually shown is that the matrix
element is the same in gauges $F(A)=w$ 
and $F(^\omega A)=w$, where
$\omega$ is an $A$-independent 
gauge transformation parameter. An outline of the proof
goes as follows (using gauge-invariant interpolating
operators for external states):
\begin{eqnarray}
&&  \langle 0|
T(J_f(x)S_g(y)J_i^\dagger(0))|0\rangle|_{F(^\omega A),G} \nonumber  \\
 &=& {1\over {\cal N}}\int [D\phi][Dw]
    \Delta_F(A) \delta(F(^\omega A)-w)G(w) J_f(\phi)
        S_g(\phi) J_i^\dagger(\phi) e^{iS} \nonumber \\
  &=&  {1\over {\cal N}}\int [D\phi][Dw]
       \Delta_F(A)\delta(F(A)-w)G(w) J_f(\phi)
        S_g(^{\omega^{-1}}\phi) J_i^\dagger(\phi) e^{iS}
\end{eqnarray}
where after the second equality one has made a gauge 
transformation $\omega^{-1}$, and all the $\omega$
dependence now appears in $S_g$. Write
\begin{equation}
    S_g(^{\omega^{-1}}\phi) \equiv S_g(\phi) + \delta S_g(\phi, \omega)\ . 
\end{equation}
Chen and Wang showed that $\delta S_g(\phi,\omega)$
has a zero matrix element in a helicity state and so, 
\begin{equation}
 \langle 0|
T(J_i(x)S_g(y)J_f^\dagger(0))|0\rangle|_{F(^\omega A),G}
  =  \langle 0|
T(J_i(x)S_g(y)J_f^\dagger(0))|0\rangle|_{F(A),G} \ .
\label{cw} 
\end{equation}
As we have demonstrated above, Eq.(\ref{cw}) is not sufficient 
to guarantee that that matrix element remains invariant
under an arbitrary new gauge condition $\tilde G(A)=\tilde w$. 
Indeed, under $A$-dependent gauge transformations, 
the proof in Ref. \cite{chen} no longer applies. 

As an example of $A$-dependent gauge transformation, 
we compare calculations in the covariant gauge and light-cone
gauge ($A^+=0$). The vacuum matrix element of a general operator
$\hat O$ (which can be a time-ordered product of operators
at several different spacetime points) in the covariant gauge 
can be written as, 
\begin{equation}
  \langle 0 |\hat O|0\rangle |_L
   = {1\over {\cal N}}
     \int [D\phi][Dw]\Delta_F(A)\delta(\partial\cdot A-w)
       O(\phi) \exp(-iw^2/2\lambda)\exp(iS)\ ,
\end{equation}
where we have included the standard Gaussian weighting
function. The same matrix element in light-cone 
gauge $A^+=0$ can be expressed as, 
\begin{equation}
  \langle 0 |\hat O|0\rangle |_A
   = {1\over {\cal N}}
     \int [D\phi]\Delta_F(A)\delta(A^+)
       O(\phi) \exp(iS) \ . 
\end{equation}
Using the same method used in deriving Eq.(9), we can show
that the matrix element in two different gauges is
related by the following equation: 
\begin{equation}
   \langle 0|\hat O(\phi)|0\rangle|_A
   = \langle 0|\hat O(^\omega\phi)|0\rangle|_L \ ,
\label{relation}
\end{equation}
where $\omega$ is a gauge transformation which brings
a gauge configuration to the $A^+=0$ gauge, 
\begin{equation}
     \omega = -{1\over D^+}A^+ \ . 
\label{trans}
\end{equation}
Equation (\ref{relation}) 
is important. It relates any matrix element
(or Green's function) in the axial gauge 
to that of a gauge-transformed 
operator in the covariant gauge!  
For instance, using the relation 
one can easily recover the gluon or
quark field propagators in the light-cone gauge 
from those in the covariant gauge. 
We emphasize here that similar relations between any two
independent gauges can be derived and they are 
entirely {\it nonperturbative}. 
The relationship between the covariant and
axial gauge calculations has also 
been studied recently by Joglekar and collaborators
\cite{joglekar}. In those studies, generalized
BRST transformations are used to connect different
gauge choices.

As an application of Eq.(\ref{relation}), we 
consider the difference of the $\vec{S}_g$
matrix elements in the covariant and light-cone 
gauges,
\begin{eqnarray}
    \langle f |S_g| i\rangle|_A -  \langle f |S_g| i\rangle|_L 
  = \langle f|S_g(^\omega A) - S_g(A)|i\rangle|_L\ , 
\label{check}
\end{eqnarray}
which can be calculated completely 
in the covariant gauge. 
Given $\omega$ in Eq.(\ref{trans}), one cannot 
show, perturbatively or nonperturbatively, 
that the right-hand side vanishes when the 
external states have definite helicity. In fact
the above equation can be used to directly check 
the one-loop calculation presented
in Ref. \cite{hoodbhoy}. 
Consider an ``on-shell'' quark in the state of momentum $p^\mu$ and
helicity 1/2. We notice that the quark fields are 
not gauge invariant. However,
the difference between the original and the 
gauge-transformed quark fields has no perturbative 
quark pole; therefore, Eq. (\ref{check}) still
applies. At one-loop 
order we find, 
\begin{equation}
    \left\langle p+{1\over 2}\left|\delta S_g(A)\right|
   p+{1\over 2}\right\rangle
 = C_F {\alpha_s\over 2\pi} {\ln \left( 
     Q^2\over \mu^2\right)} \ , 
\end{equation}
where $\delta S_g = S_g(^\omega A) - S_g(A)$, 
$C_F=(N_c^2-1)/(2N_c)$ with $N_c$ the number of colors, 
and $Q^2$ and $\mu^2$ are
the ultraviolet and infrared cutoffs, respectively.
The above result is exactly what we found in Ref. \cite{hoodbhoy}. 

We have also found that in the one-loop calculation
the $\vec{S}_g$ matrix element in the covariant 
gauge is independent of the Feynman parameter $\lambda$.
By studying the relations between matrix elements
in $\partial^\mu A_\mu = w$ and $\partial^\mu A_\mu 
= w/\sqrt{\lambda}$ gauges, we can show the 
$\lambda$-independence directly.
Notice that the different choice in $\lambda$, like 
the gauge transformations considered in Ref. 
\cite{chen}, leaves the 
Faddeev-Popov determinant invariant.  

To summarize, we demonstrated that the general gauge-independence
of a physical matrix element must be checked in the path
integral formalism through gauge-field-dependent gauge
transformations. Using this, we showed that the gluon spin
contribution to the nucleon spin is indeed gauge-dependent.
The conclusion also applies readily to other gauge-dependent
operators considered in Ref.\cite{chen}.  
We derived a general relation between matrix elements
in the covariant and axial gauges. Using the
relation, we calculated the one-loop difference of 
the gluon spin contribution to the quark spin in the
two gauges. The result confirms the explicit calculation
presented in Ref. \cite{hoodbhoy}.

\acknowledgements
The authors wish to acknowledge the support of the U.S.~National
Science Foundation under grant no. INT9820072, 
and the U.S.~Department
of Energy under grant no. DE-FG02-93ER-40762.

\end{document}